\documentclass[11pt,superscriptaddress,aps,prd,preprint]{revtex4}
\usepackage{graphicx}
\usepackage{amsmath}
\usepackage{slashed}

\newcommand{\sq}{\mathrm{sech^2}}

\begin{document}
\title{Nonanalyticity of the induced Carroll-Field-Jackiw term at finite temperature}
\author{J. F. Assun\c c\~ao}
\author{T. Mariz}
\affiliation{Instituto de F\'\i sica, Universidade Federal de Alagoas,\\ 57072-900, Macei\'o, Alagoas, Brazil}
\email{jfassuncao,tmariz@fis.ufal.br}
\author{A. Yu. Petrov}
\affiliation{Departamento de F\'{\i}sica, Universidade Federal da Para\'{\i}ba,\\
 Caixa Postal 5008, 58051-970, Jo\~ao Pessoa, Para\'{\i}ba, Brazil}
\email{petrov@fisica.ufpb.br}
\begin{abstract}
In this paper, we discuss the behavior of the Carroll-Field-Jackiw (CFJ) coefficient $k^{\mu}$ arising due to integration over massive fermions, and the modification suffered by its topological structure in the finite temperature case. Our study is based on the imaginary time formalism and summation over the Matsubara frequencies. We demonstrate that the self-energy of photon is non-analytic for the small $k^{\mu}$ limit, i.e., the static limit $(k_0=0,\vec k\rightarrow 0)$ and the long wavelength limit $(k_0\rightarrow 0,\vec k= 0)$ do not commute, while the tensorial structure of the CFJ term holds in both limits.
\end{abstract}
\maketitle
\section{Introduction}

The Lorentz and, in certain cases, CPT symmetries breaking is now treated as an important ingredient of the possible extension of the standard model \cite{Colladay:1996iz,Colladay:1998fq}. Typically, the Lorentz symmetry is broken through some additive terms that are proportional to small constant vectors or tensors (coefficients) introducing privileged directions in the space-time. Their potential applications range from quantum field theory to condensed matter physics, as occurs, for example, with the well-known Carroll-Field-Jackiw (CFJ) term \cite{Carroll:1989vb} (see \cite{Grushin} for its applications to study of Weyl semi-metals). This term represents a four-dimensional extension of the Chern-Simons (CS) term, which has gained considerable attention in three dimensions (3D) because of its relation with planar phenomena, such as the fractional quantum Hall effect and superconductivity \cite{Deser:1982, Wilczek:1982}. 

The CFJ term is a topological term violating the Lorentz and CPT symmetries, while the gauge symmetry is preserved. It is responsible to provide a topological mass for the gauge field, thus changing the dispersion relation of the photon in the vacuum \cite{Jackiw:1998js}. It is important to notice that the CFJ term can be generated dynamically through the functional integration over the Dirac fermion fields in a classical action involving a coupling of fermions to the gauge field and to a constant axial vector \cite{Jackiw:1999yp}. Therefore, we can speak about the effective Lorentz-breaking dynamics of the gauge field in four dimensions.

An important feature of the CFJ term consists in its ambiguity \cite{Jackiw:1999qq}. From the formal viewpoint this ambiguity is related to the fact that the contribution to the CFJ term is superficially divergent, however, its pole parts are mutually canceled, so that we have a some kind of the removable $\infty-\infty$ singularity. Some  values for the CFJ term are presented in \cite{Jackiw:1999yp,PerezVictoria:1999uh,Chung:1998jv,Chung:1999gg,Chen:1999ws,Bonneau:2000ai,Sitenko:2001iw,Chaichian:2000eh,Andrianov:2001zj,Bazeia:2003vt}, which differ by the regularization methods used to remove the singularity, while the topological structure of this term continues to be the same in all cases.

In this paper, we focus on the behavior of the coefficient of the induced CFJ term, when the physical system is placed in contact with a thermal bath \cite{Seminara:2001,Ebert:2004pq,Mariz:2005jh,Leite:2013pca}, in order to study its nonanalyticity. It is known that in the regime of finite temperature, the self-energy is, in many cases, a nonanalytic function of the external momenta $k_{\mu}$, in the limit $k_\mu\to 0$. Due to using of a specific frame of the thermal bath, the self-energy, in general, depends on $k_0$ and $\vec k$ in different manners, so that the static limit $(k_0=0,\vec k\rightarrow 0)$ and the long wavelength limit $(k_0\rightarrow 0,\vec k=0)$ in these cases do not commute, as it was demonstrated in the cases of self-interacting scalars \cite{Das:1997gg}, Maxwell-Chern-Simons-Higgs model \cite{Alves:2001nm}, 3D QED \cite{KaoYang:1993} and hot QCD \cite{Gross:1980br,Weldon:1982aq,Frenkel:1989br}. As the ultraviolet behavior of the theory (and hence, the structure of the possible divergences) is temperature independent, we expect that regularization ambiguities do not modify the temperature dependence of the CFJ term.

The structure of this paper is as follows. In Sec.~\ref{SE}, we single out the contribution linear in the coefficient $b_{\mu}$, from the self-energy tensor, using a perturbative method and dimensional regularization. In Sec.~\ref{CS}, we perform the computation at finite temperature, in imaginary time formalism, carrying out the sum over the Matsubara frequencies before the calculation of the spatial momentum integrals.  This calculation was never considered before, and it allows us to extract the non-covariant tensorial structure of the CFJ term, linear in the external momentum, and verify the nonanalyticity of its coefficient when we take into account the two limits $(k_0=0,\vec k\rightarrow 0)$ and $(k_0\rightarrow 0,\vec k=0)$. Finally, in Sec. IV we discuss our results and perspectives.

\section{The One-loop photon self-energy tensor}\label{SE}

In this section we are interested in studying the radiative generation of the CPT-odd Lorentz-violating Chern-Simons term from the coupling of the gauge field to massive fermions, at finite temperature. For this, let us consider the fermionic sector of the Lorentz-violating QED, given by
\begin{equation}
{\cal L}_\psi = \bar\psi(i\slashed{\partial}-m-\slashed{b}\gamma_5-e\slashed{A})\psi.
\end{equation}
Thus, the corresponding generating functional can be written as
\begin{equation}
Z[A_\mu,\bar\psi,\psi] = \int DA_{\mu}D\bar\psi D\psi\ e^{i\int d^4x{\cal L}_\psi} = \int DA_{\mu}\ e^{iS_\mathrm{eff}},
\end{equation}
where the one-loop effective action, obtained by integrating over the fermions, takes the form
\begin{equation}\label{Seff}
S_\mathrm{eff} = -i\mathrm{Tr}\ln(\slashed{p}-m-\slashed{b}\gamma_5-e\slashed{A}).
\end{equation}
Here, $\mathrm{Tr}$ stands for the trace over the Dirac matrices, as well as the trace over the space coordinate.

Now, in order to single out the quadratic terms in $A_\mu$ of the effective action, we rewrite the expression (\ref{Seff}) as
\begin{equation}
S_\mathrm{eff}=S_\mathrm{eff}^{(0)}+\sum_{n=1}^\infty S_\mathrm{eff}^{(n)},
\end{equation}
where $S_\mathrm{eff}^{(0)}=-i\mathrm{Tr}\ln(\slashed{p}-m-\slashed{b}\gamma_5)$ and 
\begin{equation}
S_\mathrm{eff}^{(n)} = \frac{i}{n}\mathrm{Tr}\left(\frac{1}{\slashed{p}-m-\slashed{b}\gamma_5}e\slashed{A}\right)^n.
\end{equation}
Then, after the evaluation of the trace over the coordinate space, by using relation $A_\mu(x) G(p)=G(p-i\partial)A_\mu(x)$, which is the basic one within the derivative expansion method \cite{Aitchison:1984ys,Aitchison:1985pp,Babu:1987rs}, and the completeness relation of the momentum space, the quadratic action $S_\mathrm{eff}^{(2)}$ is rewritten as 
\begin{equation}\label{Seff2}
S_\mathrm{eff}^{(2)} = \frac{1}{2}\int d^4x\, \Pi^{\mu\nu} A_\mu A_\nu.
\end{equation}
In the above expression $\Pi^{\mu\nu}$ is the vacuum polarization tensor, that is, in our case, the one-loop photon self-energy tensor of the form
\begin{equation}\label{Pi}
\Pi^{\mu\nu} = ie^2\int\frac{d^{4}p}{(2\pi)^4}\mathrm{tr}\,G(p)\gamma^\mu G(p-i\partial)\gamma^\nu,
\end{equation}
where the fermion propagator depends on the Lorentz-violating axial vector $b_{\mu}$ and is given by
\begin{equation}\label{G}
G(p) = \frac{1}{\slashed{p}-m-\slashed{b}\gamma_5}.
\end{equation}

In order to work out the one loop photon self-energy~(\ref{Pi}), in the perturbative approach, we must expand the propagator~(\ref{G}) up to the first order in $\slashed{b}$, as follows
\begin{equation}
G(p)=S(p)+S(p)\slashed{b}\gamma_5S(p)+\cdots,
\end{equation}
with $S(p)=(\slashed{p}-m)^{-1}$. Thus, we expand the vacuum polarization as a power series in $b^{\mu}$ as $\Pi^{\mu\nu}=\Pi_0^{\mu\nu}+\Pi_b^{\mu\nu}+\cdots$, where $\Pi_b^{\mu\nu}=\Pi_{b12|odd}^{\mu\nu}+\Pi_{b21|odd}^{\mu\nu}$, with
\begin{subequations}
\begin{eqnarray}
\Pi_{b12|odd}^{\mu\nu}(k) &=& ie^2\int\frac{d^{4}p}{(2\pi)^4}\mathrm{tr}\,S(p)\slashed{b}\gamma_5S(p)\gamma^\mu S(p_1)\gamma^\nu, \\
\Pi_{b21|odd}^{\mu\nu}(k) &=& ie^2\int\frac{d^{4}p}{(2\pi)^4}\mathrm{tr}\,S(p)\gamma^\mu S(p_1)\slashed{b}\gamma_5S(p_1)\gamma^\nu.
\end{eqnarray}
\end{subequations}
In these expressions $p_1^{\mu}=p^{\mu}-k^{\mu}$ and $k^{\mu}$ is the external momentum. Note that immediately we can see that $\Pi_{b21|odd}^{\mu\nu}(k)=\Pi_{b12|odd}^{\nu\mu}(-k)$, by considering the shift $p\to p+k$. 

Let us calculate $\Pi_{b12|odd}^{\mu\nu}(k)$ and $\Pi_{b21|odd}^{\mu\nu}(k)$, in the dimensional regularization scheme, moving the matrix $\gamma_5$ to the end of the expression, so that, after rationalizing the propagator and calculating the trace, we get
\begin{subequations}\label{Pib1221}
\begin{eqnarray}\label{Tr}
\Pi_{b12|odd}^{\mu\nu}(k) &=& 4e^2\mu^{4-D}\int\frac{d^D p}{(2\pi)^D}\frac{1}{(p^2-m^2)^2 (p_1^2-m^2)}\left[(p^2-m^2)\epsilon^{\mu\nu\kappa\lambda} b_\kappa (p_\lambda+k_\lambda)\right.\nonumber\\
&& \left.+2(p^\mu \epsilon^{\nu\kappa\lambda\rho} b_\kappa k_\lambda p_\rho-p^\nu \epsilon^{\mu\kappa\lambda\rho} b_\kappa k_\lambda p_\rho-(p\cdot k) \epsilon^{\mu\nu\kappa\lambda} b_\kappa p_\lambda)\right],\\
\label{Tr2}
\Pi_{b21|odd}^{\mu\nu}(k) &=& 4e^2\mu^{4-D}\int\frac{d^D p}{(2\pi)^D}\frac{1}{(p^2-m^2) (p_1^2-m^2)^2}\left[(p_1^2-m^2)\epsilon^{\mu\nu\kappa\lambda} b_\kappa (-p_{\lambda}+2k_\lambda)\right.\nonumber\\
&& \left.+2(p_1^\mu \epsilon^{\nu\kappa\lambda\rho} b_\kappa k_\lambda p_{\rho}-p_1^\nu \epsilon^{\mu\kappa\lambda\rho} b_\kappa k_\lambda p_{\rho}-(p_1\cdot k) \epsilon^{\mu\nu\kappa\lambda} b_\kappa p_{1\lambda})\right].
\end{eqnarray}
\end{subequations}
It is easy to see that these expressions are not explicitly gauge invariant. Indeed, linearly divergent terms in both of these expressions, that are, $(p^2-m^2)\epsilon^{\mu\nu\kappa\lambda} b_\kappa p_\lambda$ and $(p_1^2-m^2)\epsilon^{\mu\nu\kappa\lambda} b_\kappa p_\lambda$, respectively, are not transversal. However, these divergent terms do not contribute to the final result $\Pi_{b}^{\mu\nu}$, because they cancel each other already in the integrand. Thus, the remaining divergences are at most logarithmic and the gauge invariance is restored.

Nevertheless, the relation $\Pi_{b21|odd}^{\mu\nu}(k)=\Pi_{b12|odd}^{\nu\mu}(-k)$ is spoiled after the cancellation of these linear divergent terms. 
However, by considering, f.e.,
\begin{eqnarray}
\frac{(p^2-m^2)\epsilon^{\mu\nu\kappa\lambda} b_\kappa k_\lambda}{(p^2-m^2)^2(p_1^2-m^2)} = \frac{3}{2}\frac{(p^2-m^2)\epsilon^{\mu\nu\kappa\lambda} b_\kappa k_\lambda}{(p^2-m^2)^2(p_1^2-m^2)} - \frac{1}{2}\frac{(p_1^2-m^2)\epsilon^{\mu\nu\kappa\lambda} b_\kappa k_\lambda}{(p^2-m^2)(p_1^2-m^2)^2}
\end{eqnarray}
in $\Pi_{b12|odd}^{\mu\nu}(k)$, we can rewrite the expressions (\ref{Pib1221}), so that we now have $\Pi_b^{\mu\nu}=\tilde\Pi_{b12|odd}^{\mu\nu}+\tilde\Pi_{b21|odd}^{\mu\nu}$, where
\begin{eqnarray}\label{EGI}
\tilde\Pi_{b12|odd}^{\mu\nu} &=& 4e^2\mu^{4-D}\int\frac{d^D p}{(2\pi)^D}\frac{1}{(p^2-m^2)^2 (p_1^2-m^2)}\left[\frac{3}{2}(p^2-m^2)\epsilon^{\mu\nu\kappa\lambda} b_\kappa k_\lambda\right.\nonumber\\
&& \left.+2(p^\mu \epsilon^{\nu\kappa\lambda\rho} b_\kappa k_\lambda p_\rho-p^\nu \epsilon^{\mu\kappa\lambda\rho} b_\kappa k_\lambda p_\rho-(p\cdot k) \epsilon^{\mu\nu\kappa\lambda} b_\kappa p_\lambda)\right]
\end{eqnarray}
and $\tilde\Pi_{b21|odd}^{\mu\nu}(k)=\tilde\Pi_{b12|odd}^{\nu\mu}(-k)$. Therefore, we need to calculate only $\Pi_{b12|odd}^{\mu\nu}$ since these tensors contribute equally to the result $\Pi_{b}^{\mu\nu}$. We note that this expression has a structure rather typical to the integrands arising within the calculation of the CFJ term, see, f.e., \cite{Gomes:2007rv}. Also, it is clear that under the replacement $p^{\kappa}p_{\rho}\to \frac{1}{D}{g^{\kappa}}_{\rho}p^2$, either for $D=4$ or for $D=4+2\epsilon$, with $\epsilon\to 0$, the logarithmic divergence disappears, and the integral turns out finite but different for these two choices of $D$ just as occurs, f.e., in \cite{Gomes:2007rv}. So, the result is finite and ambiguous giving $\Pi_{b12|odd}^{\mu\nu}=6C\frac{e^2}{16\pi^2}\mu^{4-D}\epsilon^{\mu\nu\kappa\lambda} b_\kappa k_\lambda$, with $C$ is a regularization-dependent number yielding 1 for $D=4$ and 0 for $D=4+2\epsilon$.

The next step in the computation at finite temperature is the introduction of the Matsubara frequencies, followed by extraction of the non-covariant tensorial structure, which we perform in the next section, and, as we will see, the result will display the nonanalyticity.

\section{Nonanalyticity of the Carroll-Field-Jackiw term}\label{CS}

Let us now generalize our results to the finite temperature. To implement the finite temperature effects, we carry out the Wick rotation and split the internal momentum $p$ into its  spatial and temporal components. For this, we must perform the following replacements: $g^{\mu\nu}\to-\delta^{\mu\nu}$, i.e., $p^2\to -\delta^{\mu\nu}p^\mu p^\nu=-p^2$, $b^2\to-b^2$, $p\cdot b\to-p\cdot b$, and so on, as well as 
\begin{equation}
\mu^{4-D}\int \frac{d^D p}{(2\pi)^D} \to \mu^{3-d}\int \frac{d^d\vec p}{(2\pi)^d} \,i\int \frac{dp_0}{2\pi}
\end{equation}
and $p^\mu\to\vec p^\mu + p_0 u^\mu$, in order to separate our integration variables, with $\vec p^\mu=(0,\vec p)$, $u^\mu=(1,0,0,0)$, being a time-like vector, and $d=D-1$. In addition, let us assume from now on the system to be in thermal equilibrium with a temperature $T=\beta^{-1}$, so that the antiperiodic (periodic) boundary conditions for fermions (bosons) lead to discrete values of $p_0=(2n+1)\frac{\pi}{\beta}$ ($k_0=\frac{2\pi l}{\beta}$), where $n$ (respectively $l$) is an integer. Thus, the replacement of integration over $p_0$ by the sum over $n$ in (\ref{EGI}), by the rule ${\textstyle\int}\frac{dp_0}{2\pi}\rightarrow \frac{1}{\beta}\sum\limits_n$, gives us
\begin{eqnarray}
\Pi_{b12|odd}^{\mu\nu} &=& \frac{8ie^2}{\beta}\,\ \mu^{3-d}\sum_n \int\frac{d^d \vec p}{(2\pi)^d}\frac{p^\mu \epsilon^{\nu\kappa\lambda\rho} b_\kappa k_\lambda p_\rho-p^\nu \epsilon^{\mu\kappa\lambda\rho} b_\kappa k_\lambda p_\rho-p\cdot k \epsilon^{\mu\nu\kappa\lambda} b_\kappa p_\lambda}{(p^2+m^2)^2 (p_1^2+m^2)}\nonumber\\
&& +\frac{6ie^2}{\beta}\,\ \mu^{3-d}\sum_n \int\frac{d^d \vec p}{(2\pi)^d}\frac{\epsilon^{\mu\nu\kappa\lambda} b_\kappa k_\lambda}{(p^2+m^2) (p_1^2+m^2)}.
\end{eqnarray}
In the following, using the decomposition $p^\mu=\vec p^\mu + p_0 u^\mu$, in order to extract the tensorial structure independent of internal momentum, we arrive at
\begin{eqnarray}\label{piin}
\Pi_{b12|odd}^{\mu\nu} &=& 6ie^2\,\ \mu^{3-d}\epsilon^{\mu\nu\kappa\lambda} b_\kappa k_\lambda S_1\nonumber\\
&& +8ie^2\,\ \mu^{3-d}\left[\left(\delta^{\alpha\mu} \epsilon^{\nu\kappa\lambda\rho} b_\kappa k_\lambda \delta^\beta_\rho-\delta^{\alpha\nu} \epsilon^{\mu\kappa\lambda\rho} b_\kappa k_\lambda \delta^\beta_\rho-\vec k^\alpha \epsilon^{\mu\nu\kappa\lambda} b_\kappa \delta^\beta_\lambda\right)S_{2\alpha\beta}\right.\nonumber\\
&& +\left.\left(u^\mu \epsilon^{\nu\kappa\lambda\rho} b_\kappa k_\lambda u_\rho-u^\nu \epsilon^{\mu\kappa\lambda\rho} b_\kappa k_\lambda u_\rho-k_0 \epsilon^{\mu\nu\kappa\lambda} b_\kappa u_\lambda\right)S_3\right.\nonumber\\
&& +\left.\left(\delta^{\alpha\mu} \epsilon^{\nu\kappa\lambda\rho} b_\kappa k_\lambda u_\rho-\delta^{\alpha\nu} \epsilon^{\mu\kappa\lambda\rho} b_\kappa k_\lambda u_\rho-\vec k^\alpha \epsilon^{\mu\nu\kappa\lambda} b_\kappa u_\lambda\right)S_{4\alpha}\right.\nonumber\\
&& +\left.\left(u^\mu \epsilon^{\nu\kappa\lambda\rho} b_\kappa k_\lambda \delta^\alpha_\rho-u^\nu \epsilon^{\mu\kappa\lambda\rho} b_\kappa k_\lambda \delta^\alpha_\rho-k_0 \epsilon^{\mu\nu\kappa\lambda} b_\kappa \delta^\alpha_\lambda\right)S_{4\alpha}\right],
\end{eqnarray}
where
\begin{subequations}\label{C}
\begin{eqnarray}
\label{C1}
S_1 = \frac{1}{\beta}\sum_n\int\frac{d^d \vec p}{(2\pi)^d}\frac{1}{(p^2+m^2)(p_1^2+m^2)},\\ 
\label{C2}
S_{2\alpha\beta} = \frac{1}{\beta}\sum_n\int\frac{d^d \vec p}{(2\pi)^d}\frac{\vec p_\alpha\vec p_\beta}{(p^2+m^2)^2(p_1^2+m^2)},\\ 
\label{C3}
S_3 = \frac{1}{\beta}\sum_n\int\frac{d^d \vec p}{(2\pi)^d}\frac{p_0^2}{(p^2+m^2)^2(p_1^2+m^2)},\\ 
\label{C4}
S_{4\alpha} = \frac{1}{\beta}\sum_n\int\frac{d^d \vec p}{(2\pi)^d}\frac{p_0\vec p_\alpha}{(p^2+m^2)^2(p_1^2+m^2)}.
\end{eqnarray}
\end{subequations}
By analyzing Eq.~(\ref{piin}), we see that it involves the covariant CFJ term proportional to $S_1$ presenting as well at zero temperature and non-covariant CFJ-like structures arising only at finite temperature.

Now, let us first perform the sums and take the static $(k_0=0,\vec k\rightarrow 0)$ and long-wave $(k_0\rightarrow 0,\vec k=0)$ limits with respect to external momenta, in each one of the equations (\ref{C}), and then finally we evaluate the spatial integrals. For this, we introduce new variables, namely, $\omega_{p_1}=\sqrt{\vec{p}_1^2+m^2}$, $\omega_{p}=\sqrt{\vec p^2+m^2}$, $\xi_{p_1}=\frac{\beta\omega_{p_1}}{2\pi}$, $\xi_{p}=\frac{\beta\omega_{p}}{2\pi}$, and $\xi=\frac{\beta m}{2\pi}$. For the sake of simplicity, we also define $G=\xi_p^2+(l+i\xi_{p_1})^2$ and $\bar G=\xi_p^2+(l-i\xi_{p_1})^2$. Thus, after the sum, the coefficients (\ref{C}) take the form	
\begin{subequations}\label{D}
\begin{eqnarray}
S_1 &=& \left(\frac{\beta}{2\pi}\right)^3\int\frac{d^d \vec p}{(2\pi)^d}\frac{\xi_p\tanh(\pi\xi_{p_1})(l^2+\xi_p^2-\xi_{p_1}^2)+\xi_{p_1}\tanh(\pi\xi_p)(l^2-\xi_p^2+\xi_{p_1}^2)}{2\xi_p\xi_{p_1}[\xi_p^2+(l-i\xi_{p_1})^2][\xi_p^2+(l+i\xi_{p_1})^2]},\\
S_{2\alpha\beta} &=& \left(\frac{\beta}{2\pi}\right)^5\int\frac{d^d \vec p}{(2\pi)^d}\frac{\vec p_\alpha\vec p_\beta}{4\xi_p^3\xi_{p_1}[\xi_p^2+(l+i\xi_{p_1})^2]^2[\xi_p^2+(l-i\xi_{p_1})^2]^2}\left\{\xi_p^3\tanh(\pi\xi_{p_1})\left(G^2+\bar G^2\right)\right.\nonumber\\
&& \left. +\xi_{p_1}\tanh(\pi\xi_p)\left[8l^2 \xi_p^2(l^2+\xi_p^2+\xi_{p_1}^2)+(l^2-3\xi_p^2+\xi_{p_1}^2)(G\bar G)\right]\right.\nonumber\\
&& \left.-\pi\xi_p\xi_{p_1}\sq(\pi\xi_p)(l^2-\xi_p^2+\xi_{p_1}^2)(G\bar G)\right\},\\
S_3 &=& S_1-\left(\frac{\beta}{2\pi}\right)^5\int\frac{d^d \vec p}{(2\pi)^d}\frac{\omega_p^2}{4\xi_p^3\xi_{p_1}[\xi_p^2+(l+i\xi_{p_1})^2]^2[\xi_p^2+(l-i\xi_{p_1})^2]^2}\nonumber\\
&& \times\left\{\xi_{p_1}\tanh(\pi\xi_p)\left[8l^2 \xi_p^2(l^2+\xi_p^2+\xi_{p_1}^2)+(l^2-3\xi_p^2+\xi_{p_1}^2)(G\bar G)\right]\right.\nonumber\\
&& \left.+\xi_p^3\tanh(\pi\xi_{p_1})\left(G^2+\bar G^2\right)-\pi\xi_p\xi_{p_1}\sq(\pi\xi_p)(l^2-\xi_p^2+\xi_{p_1}^2)(G\bar G)\right\}.
\end{eqnarray}
\end{subequations}
As we can see from Eq.~(\ref{C4}), the integrand in the coefficient $S_{4\alpha}$ is odd with respect to the external momentum $(k_0,\vec k)$. So, if we choose $k_0=0\,\ (l=0)$, we have $S_4(k_0=0,\vec k)=0$, because $S_4$ is proportional to $l$. Alternatively, if we choose $\vec k= 0$, it yields a zero result, since the integrand is an odd function of internal momentum. Thus, there is no non-analyticity in this contribution.

Continuing with our analysis, for the static limit, we have
\begin{subequations}\label{SL}
\begin{eqnarray}
\label{S11}
S_1(k_0=0,\vec k\rightarrow 0) &=& \left(\frac{\beta}{2\pi}\right)^3\int\frac{d^d \vec p}{(2\pi)^d}\frac{\tanh(\pi\xi_p)-\pi\xi_p\sq(\pi\xi_p)}{4\xi_p^3},\\
\label{S21}
S_{2\alpha\beta}(k_0=0,\vec k\rightarrow 0) &=& \left(\frac{\beta}{2\pi}\right)^5\int\frac{d^d \vec p}{(2\pi)^d}\frac{\vec p_\alpha\vec p_\beta}{16\xi_p^5}\left[3\tanh(\pi\xi_p)-3\pi\xi_p\sq(\pi\xi_p)\right.\nonumber\\
&& \left.-2\pi^2\xi_p^2\tanh(\pi\xi_p)\sq(\pi\xi_p)\right],\\
\label{S31}
S_3(k_0=0,\vec k\rightarrow 0) &=& S_1(k_0=0,\vec k\rightarrow 0)-\left(\frac{\beta}{2\pi}\right)^5\int\frac{d^d \vec p}{(2\pi)^d}\frac{\omega_p^2}{16\xi_p^5}\nonumber\\
&& \times\left[3\tanh(\pi\xi_p)-3\pi\xi_p\sq(\pi\xi_p)-2\pi^2\xi_p^2\tanh(\pi\xi_p)\sq(\pi\xi_p)\right].
\end{eqnarray}
\end{subequations}
Alternatively, in the long wavelength limit, we obtain
\begin{subequations}\label{LW}
\begin{eqnarray}\label{S12}
S_1(k_0\rightarrow 0,\vec k= 0) &=& \left(\frac{\beta}{2\pi}\right)^3\int\frac{d^d \vec p}{(2\pi)^d}\frac{\tanh(\pi\xi_p)}{4\xi_p^3},\\
\label{S22}
S_{2\alpha\beta}(k_0\rightarrow 0,\vec k= 0) &=& \left(\frac{\beta}{2\pi}\right)^5\int\frac{d^d \vec p}{(2\pi)^d}\frac{\vec p_\alpha\vec p_\beta}{16\xi_p^5}\left[3\tanh(\pi\xi_p)-\pi\xi_p\sq(\pi\xi_p)\right],\\
\label{S32}
S_3(k_0\rightarrow 0,\vec k= 0) &=& S_1(k_0\rightarrow 0,\vec k= 0)-\left(\frac{\beta}{2\pi}\right)^5\int\frac{d^d \vec p}{(2\pi)^d}\frac{\omega_p^2}{16\xi_p^5}\nonumber\\
&& \times\left[3\tanh(\pi\xi_p)-\pi\xi_p\sq(\pi\xi_p)\right].
\end{eqnarray}
\end{subequations}
Therefore, as expected, in the finite temperature regime, the conditions $(k_0=0,\vec k\rightarrow 0)$ and $(k_0\rightarrow 0,\vec k= 0)$ do not commute, which means that the tensor $\Pi_{b12|odd}^{\mu\nu}$ is a nonanalytic function in the origin of momentum space, since it is not well defined at $(k_0=0,\vec{k}=0)$. It is interesting to observe that the expressions (\ref{SL}), in the static limit, look like the sum of the expressions (\ref{LW}), in the long wavelength limit, and some additive term exponentially suppressed for small temperatures, that is, at $\xi\to\infty$, but increasing as the temperature grows.

Let us now rewrite the Eq.~(\ref{piin}) in a more convenient form, by using
\begin{eqnarray}
\vec p_\alpha \vec p_\beta &\to& \frac{\vec p^2}{d} (\delta_{\alpha\beta}-u_\alpha u_\beta),
\end{eqnarray}
in coefficient $S_{2\alpha\beta}$, in order to have a tensorial structure in~(\ref{piin}) independent of the integration momentum. The result is
\begin{eqnarray}\label{SCS}
\Pi_{b12|odd}^{\mu\nu} &=& ie^2\left[\epsilon^{\mu\nu\kappa\lambda} b_\kappa k_\lambda I_1+\left(u^\mu \epsilon^{\nu\kappa\lambda\rho} b_\kappa k_\lambda u_\rho-u^\nu \epsilon^{\mu\kappa\lambda\rho} b_\kappa k_\lambda u_\rho-k_0 \epsilon^{\mu\nu\kappa\lambda} b_\kappa u_\lambda\right) I_2\right],
\end{eqnarray}
where we have defined new coefficients, given by
\begin{subequations}\label{I}
\begin{eqnarray}
\label{I1}
I_1 = 6\mu^{3-d} \left(S_1-\frac{4}{d}g^{\alpha\beta}S_{2\alpha\beta}\right),\\
\label{I2}
I_2 = 8\mu^{3-d}\left(S_3-\frac{1}{d}g^{\alpha\beta}S_{2\alpha\beta}\right).
\end{eqnarray}
\end{subequations}
It is worth to mention that the tensorial structure of~$(\ref{SCS})$ remains the same in both static and long-wave limits, unlike of other results found in literature (for example, see \cite{Brandt:2000dd}).

Our next step is to evaluate the spatial integrals of the coefficients $I_1$ and $I_2$, by using spherical coordinates in $d$ dimensions. The angular integral yields the solid angle, $\frac{2\pi^{d/2}}{\Gamma(d/2)}$, while a change of variable in radial integral from $|\vec{p}|$ to $\zeta = \frac{\beta}{2\pi}\sqrt{|\vec{p}|^2+m^2}$ will be performed. Then, for the first coefficient, by considering the static limit, we obtain
\begin{eqnarray}\label{I11}
I_1(k_0=0,\vec k\rightarrow 0) &=& \frac{3\pi^{d/2-3}(\beta\mu)^{3-d}}{16d\Gamma(d/2)}\int_{|\xi|}^{\infty}d\zeta \frac{(\zeta^2-\xi^2)^{d/2-1}}{\zeta^4}\sq(\pi\zeta)\nonumber\\
&& \times\left\{[\sinh(2\pi\zeta)-2\pi\zeta][(d-3)\zeta^2+3\xi^2]+4\pi^2\zeta^2(\zeta^2-\xi^2)\tanh(\pi\zeta)\right\},
\end{eqnarray}
which is identically zero in arbitrary dimensions. Conversely, in the inverse order of the limits, we get
\begin{eqnarray}\label{I12}
I_1(k_0\rightarrow 0,\vec k= 0) &=& \frac{3\pi^{d/2-3}(\beta\mu)^{3-d}}{16d\Gamma(d/2)}\int_{|\xi|}^{\infty}d\zeta \frac{(\zeta^2-\xi^2)^{d/2-1}}{\zeta^4}\sq(\pi\zeta)\nonumber\\
&& \times\left\{\sinh(2\pi\zeta)[(d-3)\zeta^2+3\xi^2]+2\pi\zeta(\zeta^2-\xi^2)\right\},
\end{eqnarray}
which is also identically zero in arbitrary dimensions. Therefore, the covariant term of~(\ref{SCS}) vanishes for both limits.

For the coefficient $I_2$ of the non-covariant term, we follow the same way. Then, taking into account the equations $(\ref{S21})$ and $(\ref{S31})$, in the first condition $(k_0=0,\vec k\rightarrow 0)$, we obtain
\begin{eqnarray}\label{I21}
I_2(k_0=0,\vec k\rightarrow 0) &=& \frac{\pi^{d/2-3}(\beta\mu)^{3-d}}{16d\Gamma(d/2)}\int_{|\xi|}^{\infty}d\zeta\frac{(\zeta^2-\xi^2)^{d/2-1}}{\zeta^4}\sq(\pi\zeta)\\
&&\times\left\{[\sinh(2\pi\zeta)-2\pi\zeta][(d-3)\zeta^2+3\xi^2]+4\pi^2\zeta^2[(d+1)\zeta^2-\xi^2]\tanh(\pi\zeta)\right\}.\nonumber
\end{eqnarray}
Using the expression~(\ref{I11}), we can rewrite this coefficient as
\begin{eqnarray}
I_2(k_0=0,\vec k\rightarrow 0) &=& \frac{1}{3}I_1(k_0=0,\vec k\rightarrow 0)\nonumber\\
&& +\frac{\pi^{d/2-1}(\beta\mu)^{3-d}}{4\Gamma(d/2)} \int_{|\xi|}^{\infty}d\zeta(\zeta^2-\xi^2)^{d/2-1}\sq(\pi\zeta)\tanh(\pi\zeta).
\end{eqnarray}
Note that the first term of the above equation is identically zero, while the integral in the second term is convergent, non-zero, and free of ambiguities, in the neighborhood of $d=3$. Thus, for $d=3$, we have 
\begin{eqnarray}\label{CNC1}
I_2(k_0=0,\vec k\rightarrow 0) &=& \frac{1}{2} \int_{|\xi|}^{\infty}d\zeta(\zeta^2-\xi^2)^{1/2}\sq(\pi\zeta)\tanh(\pi\zeta) = \frac{1}{2}F(\xi).
\end{eqnarray}

In the opposite limit, $(k_0\rightarrow 0,\vec k= 0)$, we must use the equations~(\ref{S22}) and~(\ref{S32}),  so that, after some changes of variables, we obtain
\begin{eqnarray}\label{I22}
I_2(k_0\rightarrow 0,\vec k= 0) &=& \frac{\pi^{d/2-3}(\beta\mu)^{3-d}}{16d\Gamma(d/2)} \int_{|\xi|}^{\infty}d\zeta\frac{(\zeta^2-\xi^2)^{d/2-1}}{\zeta^4}\sq(\pi\zeta)\nonumber\\
&& \times\left\{\sinh(2\pi\zeta)[(d-3)\zeta^2+3\xi^2]+2\pi\zeta[(d+1)\zeta^2-\xi^2]\right\}.
\end{eqnarray}
We can rewrite this coefficient also in the following way:
\begin{eqnarray}
I_2(k_0\rightarrow 0,\vec k= 0) &=& \frac{1}{3}I_1(k_0\rightarrow 0,\vec k= 0)+\frac{\pi^{d/2-2}(\beta\mu)^{3-d}}{8\Gamma(d/2)} \int_{|\xi|}^{\infty}d\zeta\frac{(\zeta^2-\xi^2)^{d/2-1}}{\zeta}\sq(\pi\zeta),
\end{eqnarray}
where we have used the coefficient~(\ref{I12}). As again the first term is identically zero and the second term is convergent, after set $d=3$, we get
\begin{eqnarray}\label{CNC2}
I_2(k_0\rightarrow 0,\vec k= 0) &=& \frac{1}{4\pi} \int_{|\xi|}^{\infty}d\zeta\frac{(\zeta^2-\xi^2)^{1/2}}{\zeta}\sq(\pi\zeta)=\frac{1}{2}G(\xi).
\end{eqnarray}

Thus, the induced CFJ term, in the regime of finite temperature, takes the form
\begin{eqnarray}\label{CSFT}
\Pi_{b}^{\mu\nu} &=& \Pi_{b12|odd}^{\mu\nu}+\Pi_{b21|odd}^{\mu\nu} = 2\Pi_{b12|odd}^{\mu\nu}\nonumber\\
&=& 2ie^2I_2\left(u^\mu \epsilon^{\nu\kappa\lambda\rho} b_\kappa k_\lambda u_\rho-u^\nu \epsilon^{\mu\kappa\lambda\rho} b_\kappa k_\lambda u_\rho-k_0 \epsilon^{\mu\nu\kappa\lambda} b_\kappa u_\lambda\right)\nonumber\\
&=& -2ie^2I_2\epsilon^{\mu\nu i\lambda} b_{i} k_\lambda.
\end{eqnarray}
As we have observed, there are different limits of $I_2$ which are described in Eqs.~(\ref{CNC1}) and (\ref{CNC2}). In addition, the resulting CFJ action, described by the linear in $b_{\mu}$ term of $(\ref{Seff2})$, is written as
\begin{eqnarray}
S_{CFJ}&=&\frac{e^2}{2}I_2\int d^4x\,\ b_{i}\epsilon^{i\lambda\mu\nu}A_{\lambda}F_{\mu\nu}.
\end{eqnarray}
As expected, the CFJ action is gauge invariant, non-covariant and nonanalytic at finite temperature. Furthermore, since the temporal component of $b_\mu$ is absent in the tensorial structure, we have only time-reversal violation, concerning the breaking of CPT symmetry.

\begin{figure}[!h]
\begin{center}\includegraphics[scale=0.8]{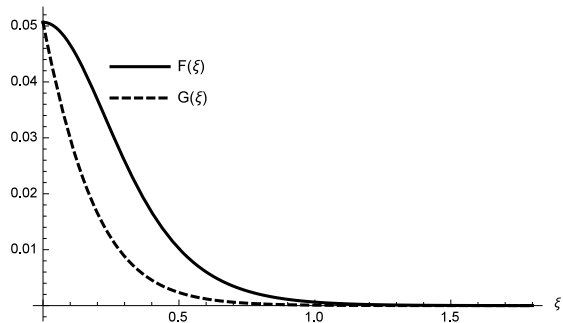}
\caption{Plot of functions $F(\xi)$ and $G(\xi)$.}
\label{FG}
\end{center}
\end{figure}

The behaviors of $F(\xi)$ and $G(\xi)$ are numerically plotted in Fig.~\ref{FG}.
 We observe that at zero and infinite temperature, the limits coincide, i.e., $\Pi_{b}^{\mu\nu}$ recovers the analyticity. When $T\rightarrow 0$ $(\xi\rightarrow \infty)$, we can easily see that (\ref{CNC1}) and (\ref{CNC2}) vanish. On the other hand, when $T\rightarrow \infty$ $(\xi\rightarrow 0)$, both $I_2(k_0=0,\vec k\rightarrow 0)$ and $I_2(k_0\rightarrow 0,\vec k= 0)$ approach $\frac{1}{4\pi^2}$, which is in agreement with the result found in \cite{Seminara:2001,Assuncao:2015lfa}, when, alternatively, $m\rightarrow 0$. In fact, the result of the static limit (\ref{CNC1})  matches the one presented in \cite{Seminara:2001}, however, the result of the long wavelength limit (\ref{CNC2}) had not yet been performed.
 
\section{Summary}

We have considered the perturbative generation of the CFJ term in the finite temperature regime. Within our study, we have considered in great details the low-energy limit of self-energy of photon in two cases, that is, the static limit and the long wavelength limit, presented in the Fig.~\ref{FG}. For this, we have performed the computation of the sum over the Matsubara frequencies before the calculation of the spatial integrals. Our result in the static limit (\ref{CNC1}) matches that one found in~\cite{Seminara:2001}, whereas the result of the long wavelength limit (\ref{CNC2}) had not yet been performed.

We have demonstrated that our result is nonanalytic, that is, the temperature dependent numerical coefficients $I_1$ and $I_2$ accompanying two different tensor structures, which by construction cannot depend on external momenta, since they represent the zero momenta limits of some functions, nevertheless, depend on the way the limits are taken. From the formal viewpoint, this nonanalyticity is a consequence of the singularity (although with no blowing-up) of the self-energy tensor, if all components of the external momentum are zero. We note, at the same time, that the massless limit in this theory is well defined being, equivalent to the high temperature limit.

A natural continuation of this study could consist in its generalization for other Lorentz-breaking contributions, for example, the aether contribution whose temperature dependence has been studied in \cite{aetherT}. We plan to do it in our next paper.

\begin{acknowledgments}
This work was supported by Conselho Nacional de Desenvolvimento Cient\'{i}fico e Tecnol\'{o}gico (CNPq) and Coordena\c{c}\~ao de Aperfei\c{c}oamento de Pessoal de N\'ivel Superior (Capes). The work
by A. Yu. P. has been supported by the CNPq project No. 303783/2015-0.
\end{acknowledgments}

\end{document}